\documentclass[conference]{IEEEtran}
\IEEEoverridecommandlockouts
\usepackage{hyperref}
\usepackage{booktabs}

\usepackage{cite}
\usepackage[pdftex]{graphicx}
\DeclareGraphicsExtensions{.pdf,.jpeg,.png}
\usepackage{amsmath}
\usepackage{comment}
\usepackage{algorithmic}
\usepackage{array}
\usepackage{fixltx2e}
\usepackage{url}
\usepackage{color,soul}
\usepackage{caption}
\usepackage{subcaption}
\usepackage{multirow}
\usepackage{authblk}
\usepackage{xcolor}
\usepackage{hyperref}

\makeatletter
\newcommand*\titleheader[1]{\gdef\@titleheader{#1}}
\AtBeginDocument{%
  \let\st@red@title\@title
  \def\@title{%
    \vspace{-0.5cm}
    \bgroup\normalfont\footnotesize\centering\@titleheader\par\egroup
    \st@red@title}
}
\makeatother

\title{Wireless On-Chip Communications for Scalable In-memory Hyperdimensional Computing}
\titleheader{This paper has been accepted at 2022 IEEE International Joint Conference on Neural Networks (IJCNN)}

\begin{document}
\bstctlcite{IEEEexample:BSTcontrol}
%\bstctlcite{IEEEexample:BSTcontrol}
%\title{Wireless Communications for an In-memory Hyperdimensional Computing Architecture}
%\title{Over-The-Air Wireless On-Chip Communications for Scalable In-memory Hyperdimensional Computing}

%\author{Robert Guirado, Abbas Rahimi, Geethan Karunaratne, Eduard Alarc\'{o}n, Abu Sebastian, Sergi Abadal%
%\thanks{R. Guirado is with Universitat Polit\'{e}cnica de Madrid (UPM), Madrid, Spain. E. Alarc\'{o}n, and S. Abadal are with Universitat Polit\`{e}cnica de Catalunya (UPC), Barcelona, Spain. A. Rahimi, G. Karunaratne, and A. Sebastian are with IBM Research Zurich GmbH, R\"{u}schlikon, Switzerland. R. Guirado was a UPC student and IBM intern during the preparation of this article.}
%}

\author[$\dagger$*]{Robert Guirado}
\author[$\dagger$]{Abbas Rahimi}
\author[$\dagger$]{Geethan Karunaratne}
\author[*]{Eduard Alarc\'on}
\author[$\dagger$]{Abu Sebastian}
\author[*]{Sergi Abadal}

\affil[$\dagger$]{IBM Research -- Zurich, R\"{u}schlikon, Switzerland}
\affil[*]{Universitat Polit\`ecnica de Catalunya, Barcelona, Spain}

\maketitle

\begin{abstract}
Hyperdimensional computing (HDC) is an emerging computing paradigm that represents, manipulates, and communicates data using very long random vectors (aka hypervectors). 
Among different hardware platforms capable of executing HDC algorithms, in-memory computing (IMC) systems have been recently proved to be one of the most energy-efficient options, due to hypervector manipulations in the memory itself that reduces data movement. Although implementations of HDC on single IMC cores have been made, their parallelization is still unresolved due to the communication challenges that these novel architectures impose and that traditional Networks-on-Chip and Networks-in-Package were not designed for. To cope with this difficulty, we propose the use of wireless on-chip communication technology in unique ways. We are particularly interested in physically distributing a large number of IMC cores performing similarity search  across a chip, and maintaining the classification accuracy when each of which is queried with a slightly different version of a bundled hypervector. To achieve it, we introduce a novel over-the-air computing that consists of defining different binary decision regions in the receivers so as to compute the logical majority operation (i.e., bundling, or superposition) required in HDC. It introduces moderate overheads of a single antenna and receiver per IMC core. By doing so, we achieve a joint broadcast distribution and computation with a performance and efficiency unattainable with wired interconnects, which in turn enables massive parallelization of the architecture.  
%(WNoC) as interconnect technology, together with a novel use of over-the-air-computing in a chip tailored to the HDC tasks, to enable the scaling of HDC architectures. 
%\hl{Our simulations show an initial parallelization potential of 64X. -- to be refined} demonstrating that the wireless approach can overcome the communication bottlenecks that arise when scaling HDC architectures, and enabling the use of a large number of IMC cores for hyperdimensional tasks, critical for its seamless parallelization.
It is demonstrated that the proposed approach allows to both bundle at least three hypervectors and scale similarity search to 64 IMC cores seamlessly, while incurring an average bit error ratio of 0.01 without any impact in the accuracy of a generic HDC-based classifier working with 512-bit vectors.
%, thanks to the exploitation of computation over the air and the relaxed error requirements of hyperdimensional computing.
\end{abstract}
\begin{comment}
\begin{IEEEkeywords}
Hyperdimensional Computing, Wireless Networks-On-Chip
\vspace{-0.3cm}
\end{IEEEkeywords}
\end{comment}

\section{Introduction} \label{introd}
\vspace{-0.1cm}

Hyperdimensional computing (HDC) is an emerging computational framework and is based on the observation that key aspects of human memory, perception and cognition can be explained by the mathematical properties of hyperdimensional spaces comprising high-dimensional vectors known as hypervectors~\cite{hdcintro}. Hypervectors are defined as $d$-dimensional (where $d \geq 1,000$) (pseudo)random vectors with independent and identically distributed components. When the dimensionality is in the thousands, a large number of quasi-orthogonal hypervectors exist. This allows HDC to combine such hypervectors into new hypervectors using well-defined vector operations, such that the resulting hypervector is unique and with the same dimension. A number of powerful computational models are built on the rich algebra of hypervectors~\cite{PlateHolographic1995,PlateHolographic2003,RachkovskijStructures2001,VSA_03}.

HDC has been employed in a range of applications such as cognitive computing~\cite{PlateAnalogy2000, SlipchenkoAnalogical2009, KanervaDollar2010}, robotics~\cite{NeubertRobotics2019}, distributed computing~\cite{VSA_Workflow,SimpkinScalable2018,TomsettDemonstrationOrch2019}, communications~\cite{CollectiveComm,Dependable_MAC_HD,Kim2018HDM,Hsu2019Collision,Hsu_HDM2,Hersche2021}, and in various aspects of machine learning. It has shown significant promise in
machine learning applications that especially demand few-shot learning~\cite{FSCIL2022,Karunaratne2021RobustHM,oneshot,RahimiBiosignal2019,RahimiEEG2017}, in-sensor adaptive learning~\cite{MoinWearable2021,BenattiEMGGestures2019}, multimodal learning~\cite{ChangEmotion2019,MitrokhinCNN2020}, and always-on smart sensing~\cite{EggimannConfigurableHD2021}. By its very nature, HDC is extremely robust in the presence of failures, defects, variations, and noise, all of which are synonymous to ultra-low energy computation. It has been shown that HDC degrades very gracefully in the presence of various faults compared to baseline classifiers: HDC tolerates  intermittent errors~\cite{HDC2016}, permanent hard errors (in memory~\cite{Li3DVRRAM2016} and logic~\cite{WuNanotube2018}), and spatio-temporal variations~\cite{HDC_NatElec20} in emerging technologies as well as noise and interference in the communication channels~\cite{Kim2018HDM,Hersche2021}. These demonstrate robust operations of HDC under low signal-to-noise ratio and high variability conditions. 
%perfectly matches with emerging hardware promising to deliver substantial energy savings, e.g., over 100$\times$ energy saving in the associative memory search compared to a digital accelerator~\cite{HDC_NatElec20}.

What these different HDC algorithms have in common is to operate on very large vectors, and therefore, are in need of architectures that handle such operations efficiently. For instance, HDC involves similarity searches across a set of stationary hypervectors in an associative memory, which are generally implemented in the form of dot-products. Due to this, in-memory computing (IMC) is a natural fit to HDC algorithms~\cite{HDC_NatElec20}. An IMC core departs from the von Neumann architectures which move data from a processing unit to a memory unit and vice versa by exploiting the possibility of performing operations (dot products, in our case) within the memory device itself~\cite{memorydevices}. This improves both the time complexity and the energy consumption of the architecture. 

IMC systems have been proposed recently to execute HDC tasks using hypervectors as wide as 10,000-bit~\cite{HDC_NatElec20}. As further elaborated in Section \ref{sec:bcg}, IMC cores are capable of computing similarity searches through dot-products with unprecedented energy-efficiency, e.g., over 100$\times$ energy saving compared to a digital accelerator~\cite{HDC_NatElec20}. However, the scaling of such architecture remains unclear due to the associated challenges. On the one hand, scaling up the architecture requires sharing a very large IMC core across many hypervectors---e.g., there will be a need to continually store and search over thousands hypervectors for representing novel classes in the incremental learning regime~\cite{FSCIL2022}---which poses a problem in terms of array impedances and programming complexity \cite{ScaleUpXbar}. On the other hand, scaling out requires deploying multiple IMC cores to execute similarity searches in parallel. This implies distribution and broadcasting hypervectors across a potentially large number of modules, which puts a large pressure on the system interconnect. 

This paper focuses on the scaling out of IMC-based HDC systems and the interconnect challenge that comes with it. In highly parallel many-core systems, Networks-on-Chip (NoC) and Networks-in-Package (NiP) are typically used to interconnect the different processing elements and ensure a correct data orchestration. However, parallelizing several similarity searches for HDC is demanding, especially when it imposes all-to-one followed by one-to-all traffic patterns, a scenario for which conventional NoCs and NiPs suffer to provide a competitive performance. Hence, the interconnect becomes a bottleneck, severely limiting the scalability of the HDC architecture.

\begin{comment}
When scaling computer architectures to many-core platforms, Networks-on-Chip (NoC) and Networks-in-Package (NiP) are typically used to interconnect the different elements and ensure a correct data orchestration. However, the communication schemes that IMC architectures require to run HDC algorithms are demanding enough so that traditional interconnect strategies might not live up to the requirements. For instance, many-to-one and one-to-many links are necessary, which make bottlenecks to appear and, in the end, limit the system performance.
\end{comment}

\begin{figure}[!t]
    \centering
    \includegraphics[width=1\columnwidth]{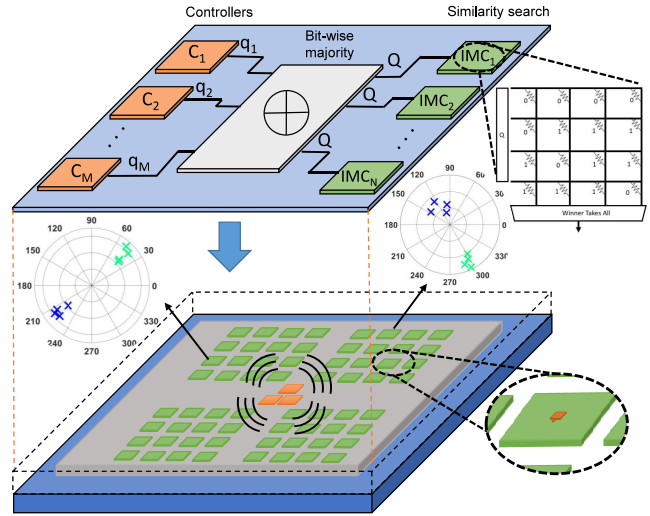}
    \vspace{-0.3cm}
    \caption{Overview of the proposed many-core wireless-enabled IMC platform. Orange encoders map to our wireless TX, while green IMCs map to our wireless-augmented IMCs. Bit-wise majority operation maps to the wireless OTA computation.}
    \label{fig:large_arch}
    \vspace{-0.5cm}
\end{figure}

To address the scalability problem of IMC-based HDC architectures, in this paper we propose to use wireless communications technology. Wireless Network-on-Chip (WNoC) have shown promise in alleviating the bottlenecks that traditional NoC and NiP face, especially for collective traffic patterns and large-scale interconnection demands that are common in HDC \cite{Laha2015, ahmed2020asymmetric, jog2021one, micro2022, wiplash}. 
%When used properly, WNoCs have a number of advantages with respect to traditional wired interconnects.
To that end, WNoCs provide native broadcast capabilities. 
%and network topology flexibility, which hold the key to opening the door to significant architectural innovations in traditional multicore processors \cite{} and deep learning accelerators \cite{}.
%or even enable some compute strategies that would just not be feasible otherwise,. 
These properties are put in use for the proposed architecture, sketched in Fig.~\ref{fig:large_arch}, with a novel approach that aims to answer the following question: \textit{\ul{Given $Q$ as a set of hypervectors that are superposed Over-The-Air (OTA), how could different physically distributed on-chip receivers reliably preform similarity search while each receiving a slightly different version of $Q$?}} To address it, we leverage the full electromagnetic knowledge of the chip package and engineer constellations to enable wireless OTA computations leading to a lightweight all-to-all concurrent communications at the chip scale. The resulting WNoC will be uniquely suited to the communication requirements of HDC operations while opportunistically bypassing the main limitations of wireless technology: the impact of relatively low aggregate bandwidth and high error rate are minimal thanks to the OTA approach and the inherent resilience of HDC algorithms to noise.

%\textcolor{red}{This way}, we are capable of parallelizing HDC tasks by leveraging the wireless broadcast capabilities, which   %communication-scalability trade-off starts to get cumbersome. 

% This enables the use of a large number of cores for hyperdimensional tasks, which in turn allows its seamless parallelization.
%by means of an electromagnetic package characterization and the unique introduction of Over-The-Air (OTA) computations in the wireless chip-scale communications scenario. 

%However, its wireless nature within a package can also become a challenge, as nodes must share a limited number of channels and, even if that is done efficiently, the aggregate bandwidth of the WNoC is generally lower than that of wired alternatives. 

This paper makes the following three novel contributions. (i) For the first time, we use a wireless interconnect solution for HDC platform that allows scaling-out similarly search across multiple independent on-chip receiver modules. (ii) For the first time too, we enable more than one simultaneous transmitter to make use of OTA computation on a chip. (iii) We leverage a pre-characterization of the chip package to optimize OTA from multiple transmitters to multiple receivers. The proposed architecture is designed and evaluated at the electromagnetic level, demonstrating that it can support up to 64 receivers with 3 transmitters with an average bit error ratio (BER) of 0.01 and the maximum BER of 0.1, which do not have any impact in the accuracy of a generic HDC-based classifier operating with 512-bit hypervectors.

The rest of the paper is organized as follows. In Sec. \ref{sec:bcg}, we provide background on the topics of HDC, IMC, and wireless communications at the chip scale. In Sec. \ref{motiv}, we motivate the problem by illustrating the scale-out of IMC-based HDC architectures and then propose the wireless solution. In Sec. \ref{meth}, we depict the simulation methodology encompassing electromagnetic simulation, signal processing, and HDC-based learning. In Sec. \ref{results}, we show the main results of the analysis. The paper is concluded in Sec. \ref{cncl}.

\section{Background}\label{sec:bcg}

\subsection{Hyperdimensional Computing}
Here we focus on a variant of HDC models by making use of pseudo-random binary vectors of thousands of dimensions~\cite{hdcintro}. When using these binary hypervectors, it is easy to find nearly unlimited non-coincident quasi-orthogonal vectors with normalized Hamming distance close to 0.5. We call these random hypervectors atomic hypervectors. In classification tasks, one can further create an encoder to operate on these atomic hypervectors by binding, bundling (i.e., superposition), and permutation operations to obtain a composite hypervector describing an object or event of interest. The composite hypervectors, generated from various examples of the same class, can be further bundled together to create a single prototype hypervector representing a class. Particularly, the bundling operation for binary hypervectors is implemented as a logical bit-wise majority operation. The prototype hypervectors are stored in the associative memory.

In the inference stage, the query hypervectors of unknown objects/events are generated by following the same procedure as in the training stage. A query hypervector is later compared to the prototype hypervectors in the associative memory. Then, the chosen label is the one assigned to the prototype hypervector that has the highest similarity to the query vector. The robustness to failure is given by the spreading of information across thousands of dimensions. See~\cite{RahimiBiosignal2019} for more details. 

%A typical HDC-based classifier works in the following way: first, in the training stage, we need to obtain prototype hypervectors representing the different classes. In order to do so, base hypervectors are assigned to describe atomic concepts in our training set, and an encoder combines the base hypervectors to generate a composite hypervector. The composite hypervectors of each class are bundled together, in order to obtain a single prototype hypervector representing the whole set. Although the bundling operation can be implemented in a number of ways, a logical (bit-wise) majority operation is typically used. In the inference stage, we find query hypervectors of unknown objects following the same methodology as in the training stage, to later compare them to the prototype hypervectors. Then, the chosen label will be the one assigned to the prototype hypervector most similar to the query vector. 

\subsection{In-memory Computing}
IMC is a non von Neumann architecture that leverages the memory unit to perform in-place computational tasks, reducing the amount of data movement and therefore cutting down the latency and energy consumption associated with in-package communication \cite{memorydevices}. That is, instead of fetching the data from the memory to the processing unit in order to carry out computations and store the results back to the memory, in IMC systems the operation is directly carried out in the computational memory, which requires less communication. 

The latency produced by memory accesses is problematic in computing systems in general, but it can be more or less harmful depending on the particular application being executed, as it can limit the overall performance of the system. When this happens, and the memory accesses become the bottleneck, the term memory wall is commonly used, referring to the disparity between the processing speed and the ability of the memory to provide data to, or receive data from, the processing units. Several memory and architecture concepts have been designed and manufactured in the recent years to overcome these problems, such as high-bandwidth memory \cite{hbm}, 2.5D and 3D monolithic integration \cite{3d}, interposers or hybrid memory cube \cite{hmc}. However, from a complete architectural point of view, these are ad-hoc solutions that are not expected to solve the problem from the root, as the fundamental problem of moving large quantities of data from memory and back remains. Instead, the novel approach of IMC is being developed and appears as a promising candidate to overcome these
challenges \cite{memorydevices}. %\RG{We might want to cut this down.}

Resistance-based IMC cores, and more specifically those based on phase-change memory (PCM) devices, have recently shown promising results~\cite{Y2022khaddamJSSC}. In a resistance-based IMC core, we can encode certain values as conductances of PCM devices placed in a mesh-like array. Then, by Ohm's law and Kirschhoff's law, a matrix-vector multiplication (MVM), essential to execute any machine learning algorithm, is as simple as tuning conductances to match the matrix values, inputting the vector as voltages from one side and finally reading the output currents from a perpendicular side.

\begin{figure} [!htb]
    \centering
    \includegraphics[width=0.9\columnwidth]{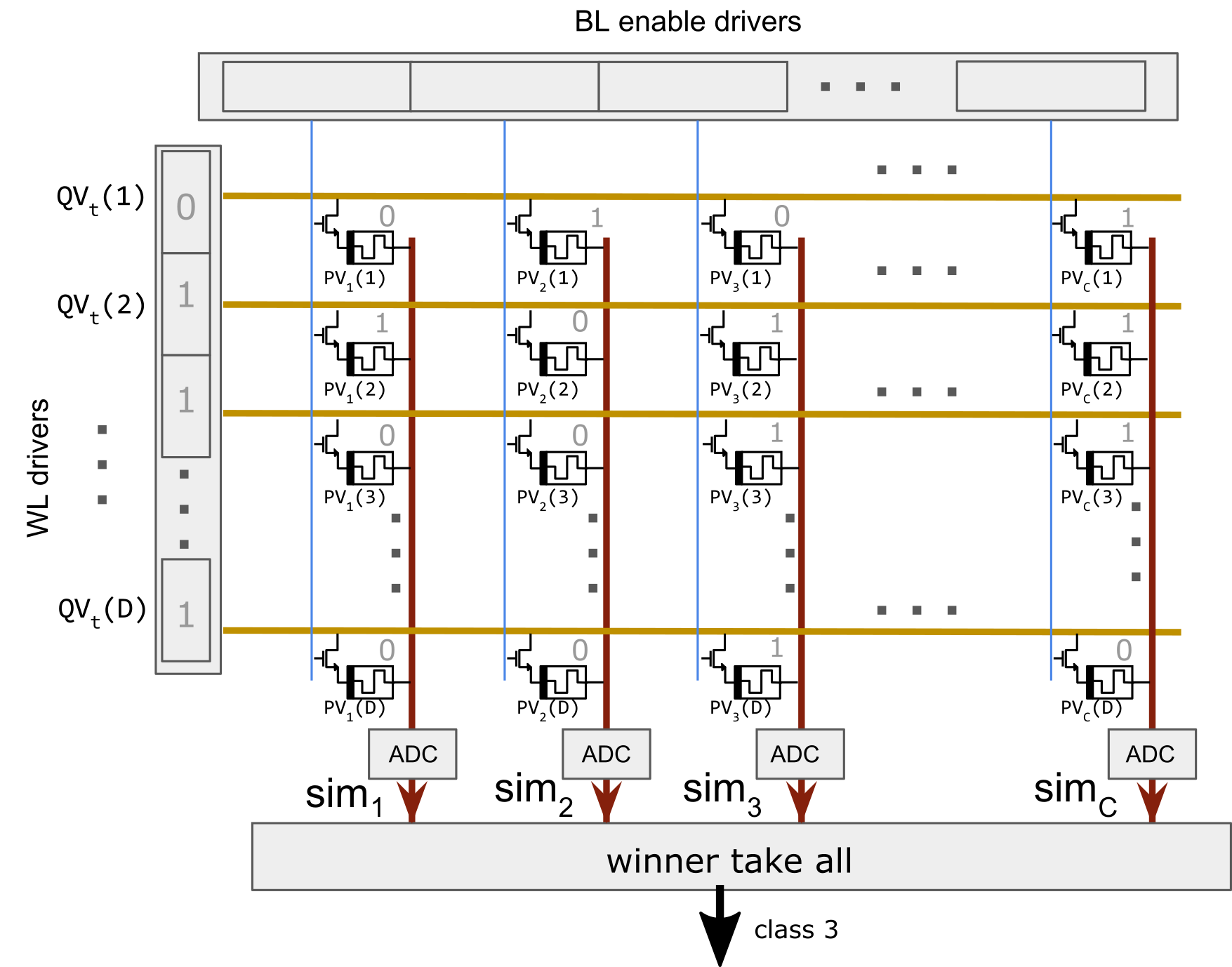}
    \vspace{-0.1cm}
    \caption{Similarity search example in an IMC core. Since the prototype hypervector of the third column is the most similar one to the query vector $Q$, it will output more current than the others and its associated label will be chosen.} %\RG{Zoom in a resistor showing the device response}}
    \label{fig:dotp}
    \vspace{-0.3cm}
\end{figure}

Although IMC architectures are capable of executing various HDC operations~\cite{HDC_NatElec20}, we are particularly interested in the similarity search in the associative memory. As shown in Fig. \ref{fig:dotp}, since the prototype hypervectors $P_i$ will be programmed in an IMC core, the similarity search through the dot product can be implemented as a MVM with the query hypervector $Q$ as input vector. This allows performing a dot-product in $O(1)$ time complexity.

\subsection{Wireless Network-on-Chip}
%\textcolor{red}{mention time-invariance of channel in the package?}
NoCs are currently the \emph{de facto} standard interconnects in modern multiprocessors due to their low latency and high throughput capabilities in systems with a few dozen processing cores. However, NoCs face significant challenges when scaling the architectures or when facing specific communication patterns such as broadcast or reductions. This has led to the point where systems are starting to be communication-bounded instead of computation-bounded. WNoCs have been introduced, among other alternatives, to overcome these issues. WNoCs are the result of augmenting cores or groups of cores with RF transceivers and antennas allowing them to communicate wirelessly through the chip package with all cores that are within range \cite{barrier, timoneda,wienna}. Even though this technology is still under development, proof-of-concept designs have been successfully implemented and tested \cite{multichannel}. 

Among the key advantages of WNoCs, one can find a natural support to broadcast communications, reduced latency, and an adaptive network topology \cite{ahmed2020asymmetric,adaptive, wiplash, imani2022smart}. Hence, WNoCs can be especially advantageous if they are used to serve specific communication patterns that are very challenging to tackle using conventional NoCs \cite{wienna}. This is of relevance in this work, as HDC algorithms being executed over IMC platforms make an intensive use of broadcast and reduction patterns, leading to important bottlenecks when scaled over traditional NoC/NiP platforms. In this case, the key strength of WNoCs lies on its use for broadcast communication, while it is in principle less suited to all-to-one reduction patterns. However, as we detail next, thanks to the proposed OTA computing solution, WNoCs become a perfect candidate to enabling the scalability of IMC-based HDC architectures.

\begin{comment}
\begin{figure} [!htb]
    \centering
    \vspace{-0.4cm}
    \includegraphics[height=3cm,width=\columnwidth]{images/Fig1.pdf}
    \caption{caption.}
    
    \label{image}
    \vspace{-0.3cm}
\end{figure}
\end{comment}

\section{Towards Wireless-Enabled Scale-Out\\ HDC Architectures} \label{motiv}

\begin{comment}
\textcolor{red}{
Centrar en hdc, explicar imc com un enabler i explicar el bottleneck. Com sescala a dreta i esquerra. Començar com que HDC pot començar trencant de veritat. Aquesta no es la.unica solucio, pero si q remarcar q es first time ever.
Mes controllers --> mes theoughput de.computacio
Anticipating a future bottleneck....
FIg 6.18 de la tesi motivation}
\end{comment}
Although HDC has a great potential and IMC systems are used to execute it efficiently, the scaling of such systems, as essential as it is to satisfy the insatiable appetite of machine learning for computational resources, is still a pending matter. In architectural terms, IMC-based HDC systems can be scaled by either increasing the size of the IMC cores (scale-up) or by placing more cores in the system (scale-out). 
%The former scaling strategy is especially troublesome, and the reason for this is twofold: on the one hand, the interconnect resistance increases as the in-memory array becomes larger. This can become a problem when the resistance becomes comparable to the device input impedance, as currents would not behave as expected \cite{}. On the other hand, by increasing the array size, the control complexity rapidly increases, not only for running MVMs but also for the weight programming mechanisms that need to sweep each diagonal in the array \cite{}. 

%However, scaling them is essential 
%parallelize computations by, for instance, performing multiple similarity searches simultaneously and hence increasing the overall computational throughput of the system. Without scaling the system, this would lead to MVMs larger than the PCM array. 
%In this case, we would have to re-program the devices and run partial MVMs several times using subsets of the prototype hypervectors, which leads to a significant increase in latency. 

\begin{figure*}[!t]
    \centering
    \begin{subfigure}{0.38\textwidth}
    \includegraphics[width=\textwidth]{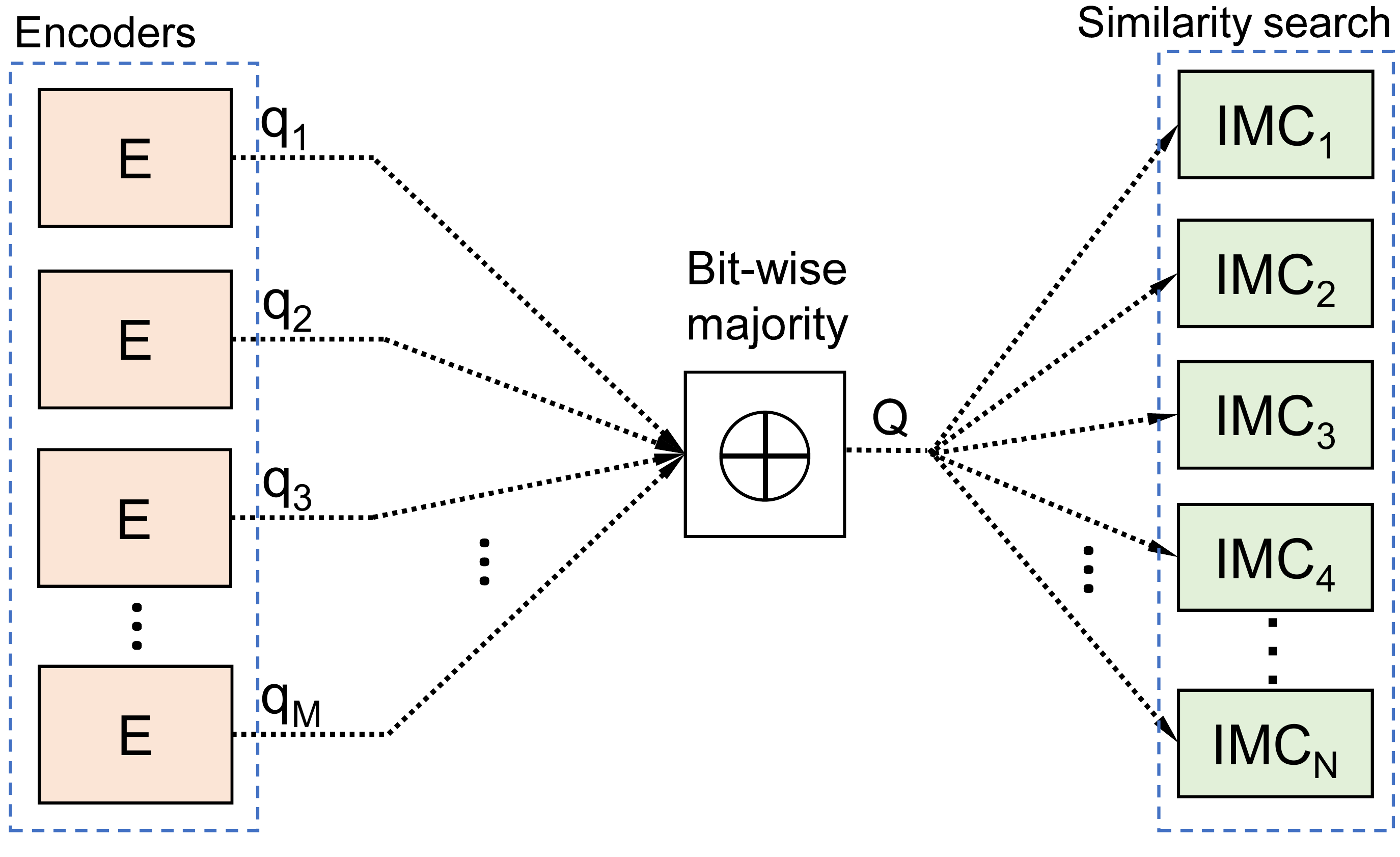}
    \caption{Logical view.}
    \label{fig:diagram}
    \end{subfigure} \hfill
    \begin{subfigure}{0.6\textwidth} \vspace{0.6cm}
    \includegraphics[width=\textwidth]{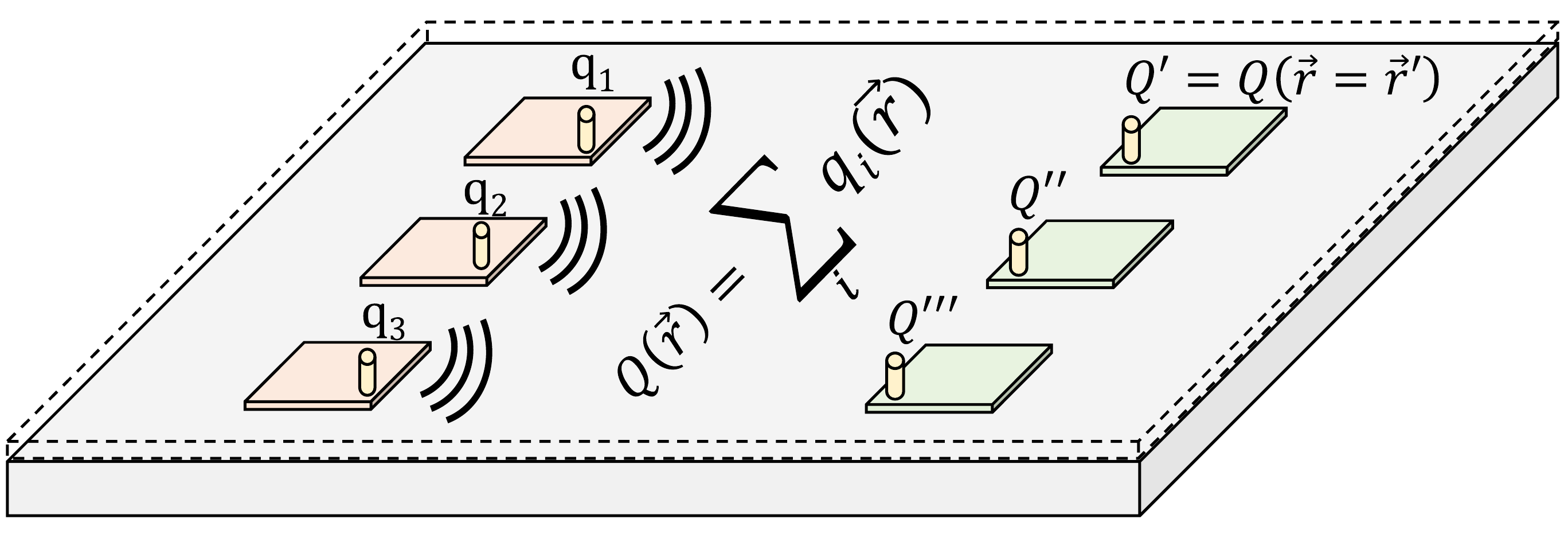}
    \caption{Wireless-enabled realization with OTA computing.}
     \label{fig:wireless_arch_diagram}
    \end{subfigure}
    \vspace{-0.1cm}
    \caption{Proposed scale-out approach of a HDC platform involving $M$ encoders generating queries $q_{1}\cdots q_{M}$, the computation of a composite query $Q$ via bit-wise majority, and $N$ IMC cores performing similarity search over multiple copies of $Q$. In the wireless case, the IMC cores receive different versions of $Q$ ($Q'$, $Q''$, $Q'''$) that are decoded minimizing the distance to $Q$.}
    \vspace{-0.3cm}
\end{figure*}

% \begin{figure}[!t]
%     \centering
%     %\includegraphics[width=0.7\columnwidth]{images/wired_alternative.png}
%     \includegraphics[width=0.9\columnwidth]{images/arch_logic_v3.pdf}
%     \vspace{-0.2cm}
%     \caption{Logical diagram of a HDC platform with $M$ encoders generating queries $q_{1}\cdots q_{M}$, a centralized core computing a composite query $Q$ through bit-wise majority, and $N$ IMC cores performing similarity search over $Q$.}
%     \label{fig:diagram}
%     \vspace{-0.3cm}
% \end{figure}

% \begin{figure}[!t]
%      \centering
%      \vspace{-0.1cm}
%      \includegraphics[width=1\columnwidth]{images/fig_wireless_artist3.pdf}
%      \caption{Wireless HDC architecture example. Each encoder sends its query hypervector $q_i$ through a modulated signal with a certain phase. The signals add up to $Q$, but each IMC receives a different version of $Q$ namely $Q'$, $Q''$, $Q'''$.}
%      \label{fig:wireless_arch_diagram}
%      \vspace{-0.3cm}
%  \end{figure}

On the one hand, scaling-up becomes complex as the required in-memory wire length blows up exponentially with the array size, leading to issues related to wire resistance and parasitic effects. Moreover, the complexity of weight programming also increases with the array size~\cite{ScaleUpXbar}. 

On the other hand, scaling-out is a technologically viable alternative. Fig.~\ref{fig:diagram} shows a logical diagram of the desired scaled-out IMC architecture, capable of executing a HDC-based classifier. The $M$ encoders at the left compute the different query hypervectors, which will be bundled later on through the majority operation. Each encoder can encode data from e.g., different sensory modalities~\cite{ChangEmotion2019,MitrokhinCNN2020}, or streaming channels~\cite{Hersche2021}. This is highly desirable since by doing a bundling of $M$ queries, we virtually increase the throughput by a factor of $M$. That is, we compress all the queries information in a single one instead of having $M$ independent transmissions and redundant bundling at the similarity search cores. The $N$ IMC cores, at the right of Fig.~\ref{fig:diagram}, are in charge of comparing the composite query hypervector with all the prototype hypervectors they have stored, enabling the aforementioned scaling-out. By following this modular approach, a system as powerful as required by each application could be designed by varying $M$ and $N$.

\vspace{0.1cm} \noindent
\textbf{Challenges of wired scale-out.}
Notwithstanding, scaling out casts a significant pressure to the system interconnect. Firstly, the interconnection between the $M$ encoders and a hypothetical circuit performing the bit-wise majority would result in heavy reduction $M$-to-1 traffic. Should the bundling operation be performed using a wired interconnect, we would have to add a centralized processing core with extra circuitry, which would not scale linearly with the number of encoders. Secondly, the interconnection between the bundling block and the $N$ IMC cores follows a broadcast topology, which becomes slow and inefficient as $N$ grows \cite{ahmed2020asymmetric}. 

%Therefore, even though the benefits of such parallelization are apparent, scaling the architecture to a growing number of cores by means of a wired fabric is challenging. %, since the needed interconnect would be too large as it would not scale in size nor in control complexity.
Even in the case of full co-integration of the encoders with specialized bundling circuitry and IMC cores, the system would need to provision a non-scalable amount resources. A lower cost modular alternative, proposed in other deep learning acceleration systems \cite{simba}, is to build the architecture with specialized chiplets and to integrate them through an interposer. In this case, however, the interposer becomes a bottleneck in terms of bandwidth and connectivity due to I/O pin limitations. This leads to multi-hop and serial-link schemes that add significant energy and latency per hop, i.e., $\sim$1 pJ and $\sim$20 ns \cite{simba}, with hop counts typically scaling with $\sqrt{N}$ for unicasts and with $N$ for broadcasts \cite{wienna}. 

In summary, wired scale-out of HDC platforms is challenging because: (i) the reduction (all-to-one) pattern generated by the bundling operation not only creates a communication bottleneck, but also acts as an implicit barrier; (ii) the broadcast (one-to-all) pattern of query distribution is inherently costly in chiplet-based systems; and (iii) both operations are sequential.

%Moreover, the multi-hop connectivity, which might be needed in the IMC cores side, could limit the performance becoming a bottleneck.  

\begin{figure}[!t]
    \centering
    \includegraphics[width=1\columnwidth]{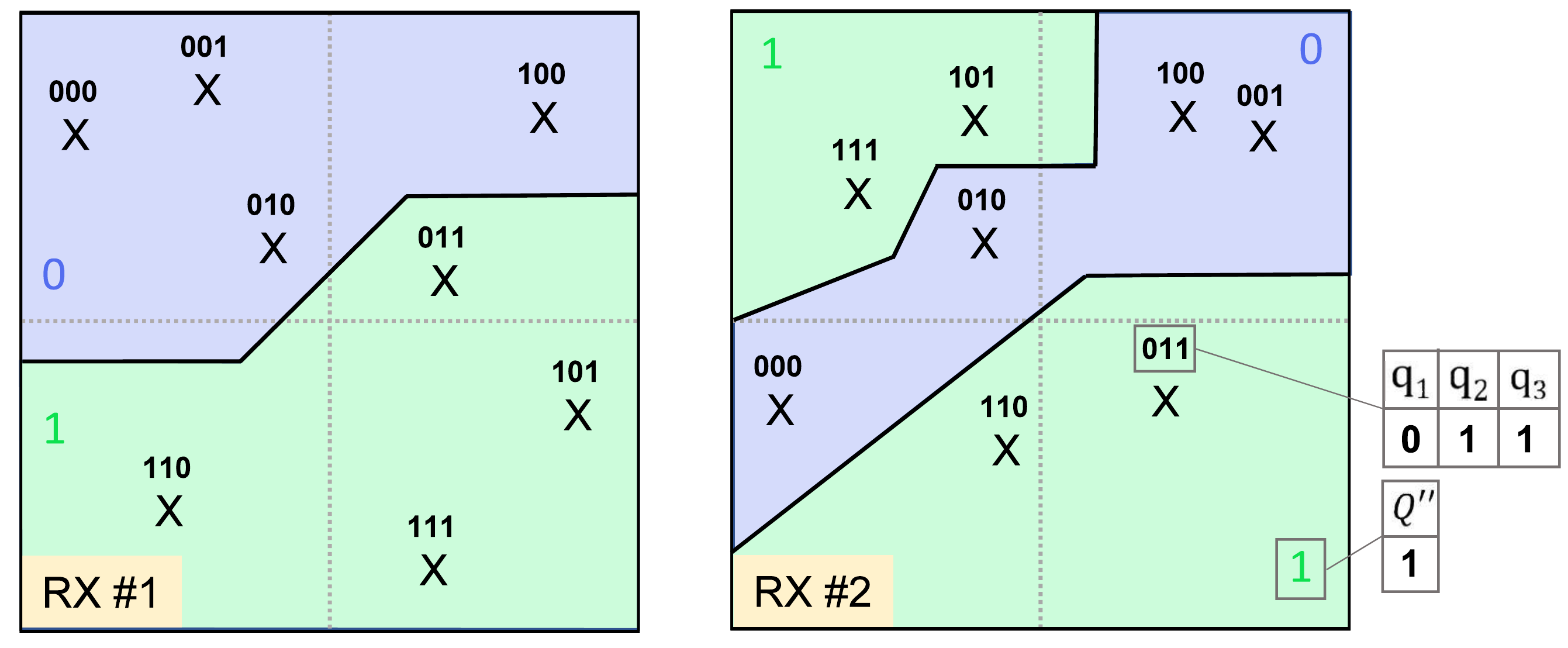}
    \vspace{-0.1cm}
    \caption{Example of decision regions of over-the-air (OTA) majority computation for three transmitters $\{q_1, q_2, q_3\}$ at two distinct receivers. Blue/green regions map to 0/1.}
    \label{fig:regions_ex}
    \vspace{-0.3cm}
\end{figure}

\noindent
\textbf{Proposed architecture.} We tackle the three problems of wired scale-out at once by augmenting a many-core HDC platform with a WNoC. Fig.~\ref{fig:wireless_arch_diagram} shows the proposed WNoC implementation with $M$ encoders augmented with wireless TXs and $N$ IMC cores augmented with wireless RXs. The encoders broadcast, in a concurrent fashion and using a single channel, the different queries to be bundled. As a result of the wave propagation, each receiver will obtain a slightly different version of the superposition of all transmitted signals, which will be decoded using the channel state information, which is quasi-static and known a priori. 
%By knowing the medium effect in each of the superposed signals, 
Hence, the final majority result is known in the RXs per each TX bit combination. That is, we can pre-assign different decision regions that map the received superposed symbols to their logical majority per each RX, as illustrated in Fig. \ref{fig:regions_ex}. See Sec.~\ref{meth} for more details. 

\begin{figure*}[!t]
    \centering
    \includegraphics[width=0.53\textwidth]{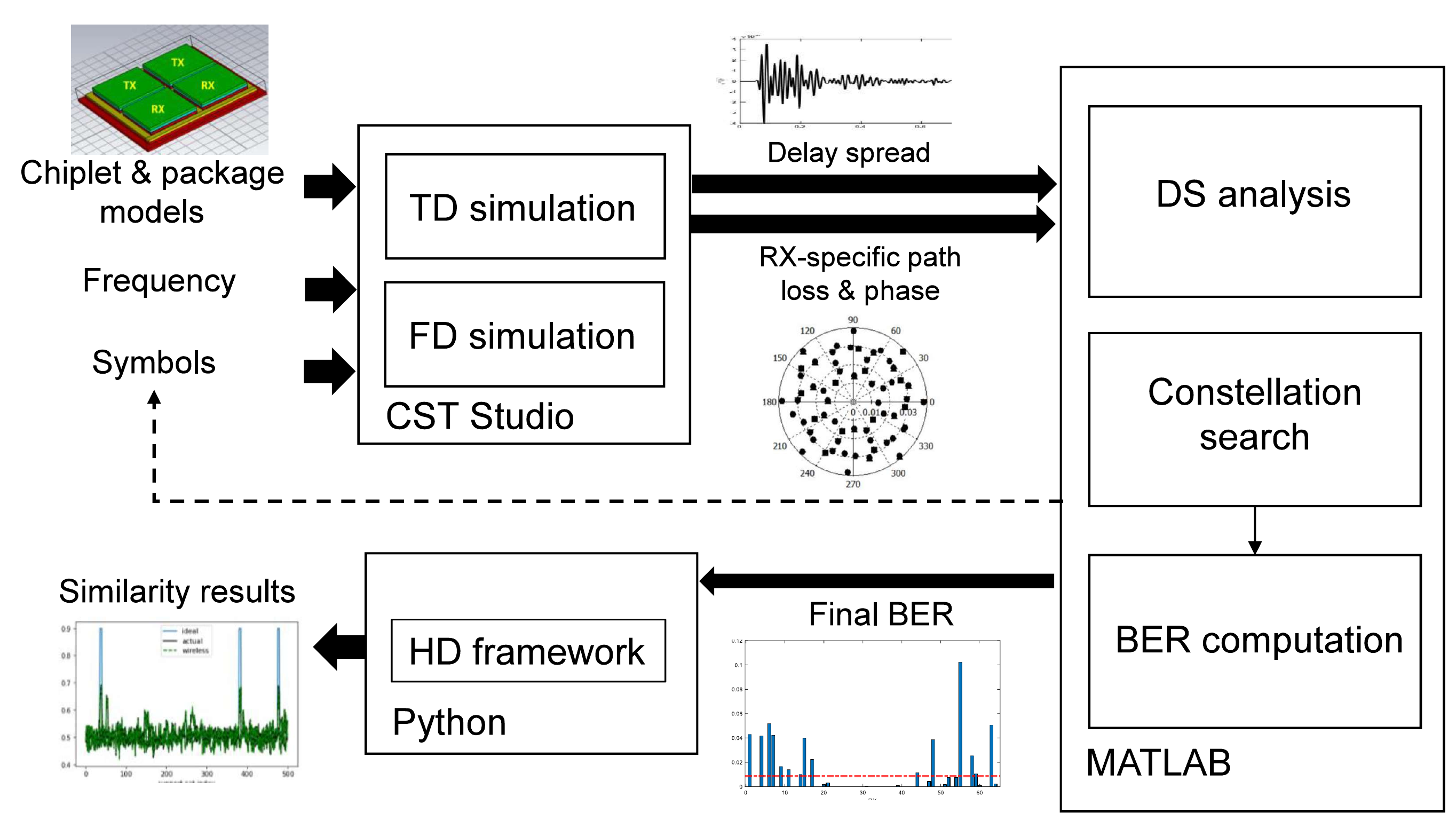}\hfill
    \includegraphics[width=0.47\textwidth]{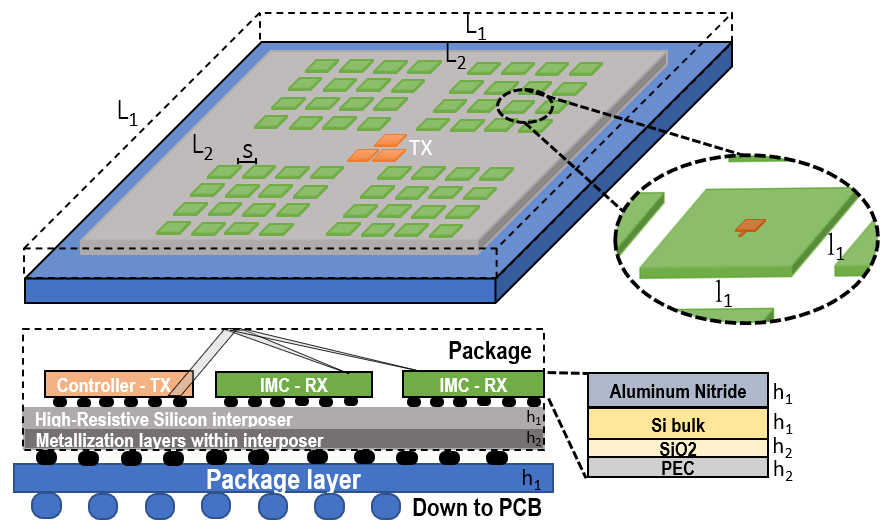}
    \vspace{-0.4cm}
    \caption{Overview of the evaluation methodology and layout of a sample architecture with 3 TXs and 64 RXs. The package is enclosed in a metallic lid and empty spaces are filled with vacuum. $h_1=0.1$ mm; $h_2 = 0.01$ mm; $l_1 = 7.5$ mm; $s = 3.75$ mm; $L_1 = 33$ mm; $L_2 = 30$ mm.}
    \label{fig:meth}\label{fig:arch}
    \vspace{-0.5cm}
\end{figure*}

In summary, the proposed architecture is built upon three key observations:
\begin{itemize}
    \item \emph{Given the controlled package scenario, OTA computing can be leveraged.} In particular, the majority operations required by the bundling of hypervectors can be performed over-the-air (OTA) with low error thanks to a pre-characterization of the channel.
    \item \emph{The inherent broadcast nature of wireless communication allows to implement single-hop in-package transfers.} This, together with the OTA bundling, allows for a seamless parallelization of the similarity search over multiple associative memories at the chip scale while completely eliminating the communication bottleneck.
    \item \emph{The resilience of the HDC paradigm to errors makes it highly tolerant to poor BER conditions}. Indeed, a drawback of wireless technology in general and OTA computing in particular is that it can suffer from relatively high error rates, leading to inefficient designs. %is generally less efficient than wires, meaning that it suffers higher error rates for the same power and link length. 
    However, as we show later in the paper, HDC is inherently resistant to such conditions and allows to scale the proposed approach to tens of IMC cores. % errors as we show later  we show that up to a BER in the order of 0.1, the HDC classification task is virtually unaffected by the errors in the communications channel. 
\end{itemize}

\section{Methodology} \label{meth}
The main contribution of this work is the validation of the OTA on-chip computing concept and scalability assuming a realistic chip package. Fig. \ref{fig:meth} summarizes the procedures followed to evaluate the proposed approach. First, a package has been modelled in CST Studio \cite{cst} together with its corresponding chiplets, as also shown in Fig. \ref{fig:arch}. The operating frequency is 60 GHz, compatible with the on-chip environment \cite{timoneda}. Symbols are transmitted with an amplitude of 0 dBm per antenna \cite{multichannel}, and the phase is discretized in 45 degree steps. Both time-domain and frequency-domain simulations for a simultaneous excitation of all TXs have been performed. The results have been post-processed to extract delay spread, path-loss data and phase data. Next, this has been used in MATLAB to perform a constellation search. That is, among all the different possible symbol phases and for all TX bit-combinations, the ones reporting the best BERs have been chosen. Finally, the error rate figures have been used in an HDC framework in order to characterize the impact of the wireless channel in the overall architecture in terms of classification accuracy.

% \begin{figure}[!t]
%     \centering
%     \vspace{-0.1cm}
%     \includegraphics[width=1\columnwidth]{images/methodology_figure.pdf}
%     \caption{Methodology overview.}
%     \label{fig:meth}
%     \vspace{-0.3cm}
% \end{figure}

% \begin{figure}[!t]
%     \centering
%     %\includegraphics[width=0.7\columnwidth]{images/arch_64.png}
%     \includegraphics[width=1\columnwidth]{images/aerial_crosssection_V4.png}
%     \vspace{-0.1cm}
%     \caption{Layout of a sample architecture with 3 TXs and 64 RXs. The package is enclosed with a metallic lid and empty spaces are filled with vacuum. $h_1=0.1$ mm; $h_2 = 0.01$ mm; $l_1 = 7.5$ mm; $s = 3.75$ mm; $L_1 = 33$ mm; $L_2 = 30$ mm.}%Immediately on the Aluminum Nitride layer, and at the ends of the package, everything is metallically encapsulated.
%     \label{fig:arch} 
%     \vspace{-0.3cm}
% \end{figure}

\noindent
\textbf{Source coding.} The way the TX encode the bits of their queries is by varying their phases. That is, all TX symbols will have same amplitude but different phases. We sweep a discrete set of 8 phases in the TXs in order to characterize the electromagnetic behaviour in each case and to find the best separable phase combinations. That is, we consider as RX constellation the aggregation of all the possible TX combinations. When choosing the optimal TX phases (two per sender, each one assigned to the binary 1 or 0), however, we have two points to consider: first, we have to meet the independent phase requirement. That is, we have to make sure that each TX only uses two phases and that the phase of each TX is independent of each other; secondly, the TX phases affect all RXs, meaning that, when we fix the symbol phases we fix the received constellation for all receivers. This implies that a joint optimization considering all RXs is needed.

\begin{figure}[!t]
    \centering
    %\begin{subfigure}{0.9\columnwidth}
    \includegraphics[width=0.9\columnwidth]{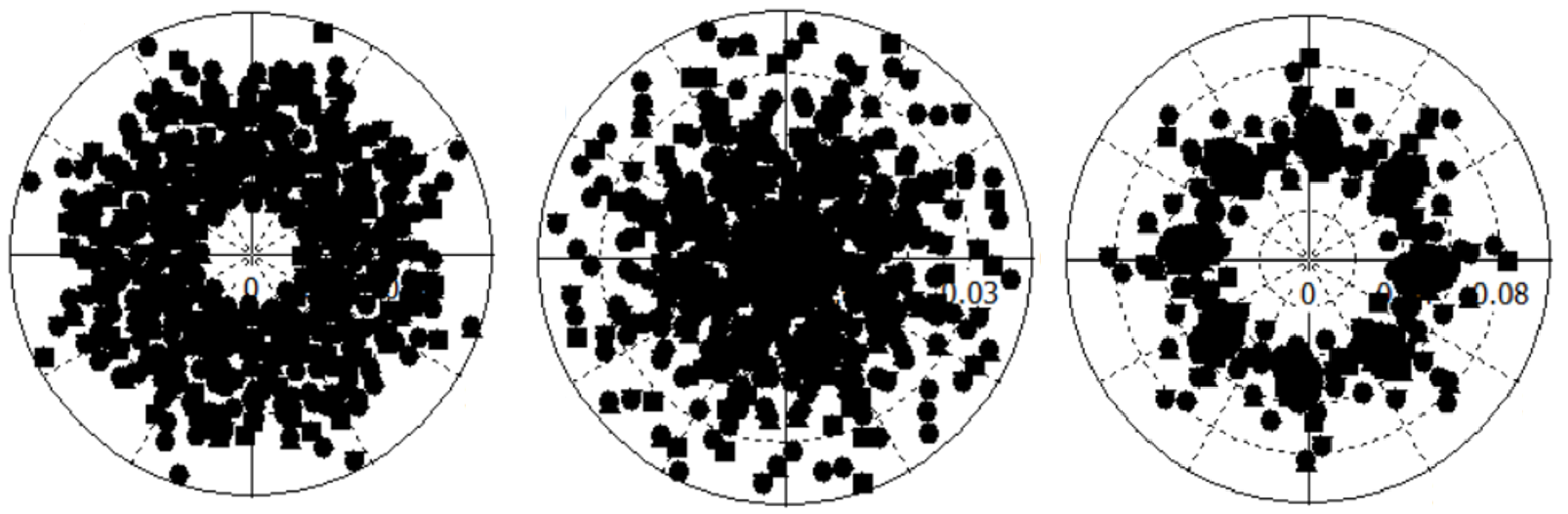}
    %\caption{Electromagnetic phase sweep outcome}
    %\end{subfigure}
    %\begin{subfigure}{0.9\columnwidth}
    \includegraphics[width=0.9\columnwidth]{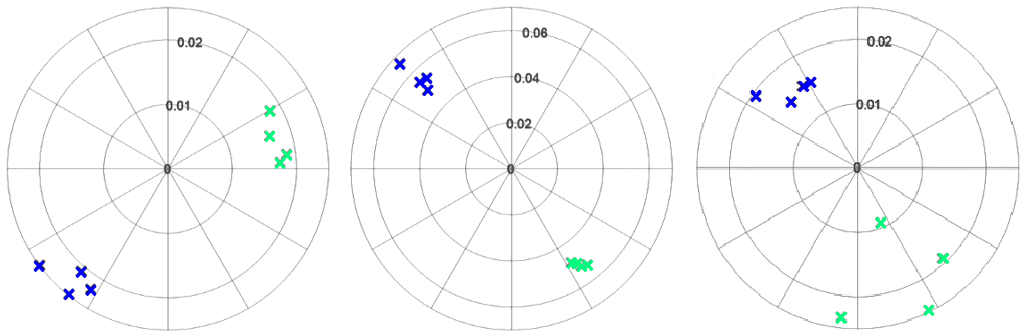}
    %\caption{Chosen constellations. Blue symbols map to logical 0 and green symbols map to logical 1.}
    %\end{subfigure}
    \caption{Sweep of all possible phase combinations (top) and chosen to minimize the error rate of the majority computation (bottom). Blue/green symbols map to logical 0/1.}
    \label{fig:ota_constellations}\label{fig:cst_res}
\end{figure}

\begin{figure}[!t]
    \centering
    \vspace{-0.3cm}
    \includegraphics[width=0.8\columnwidth]{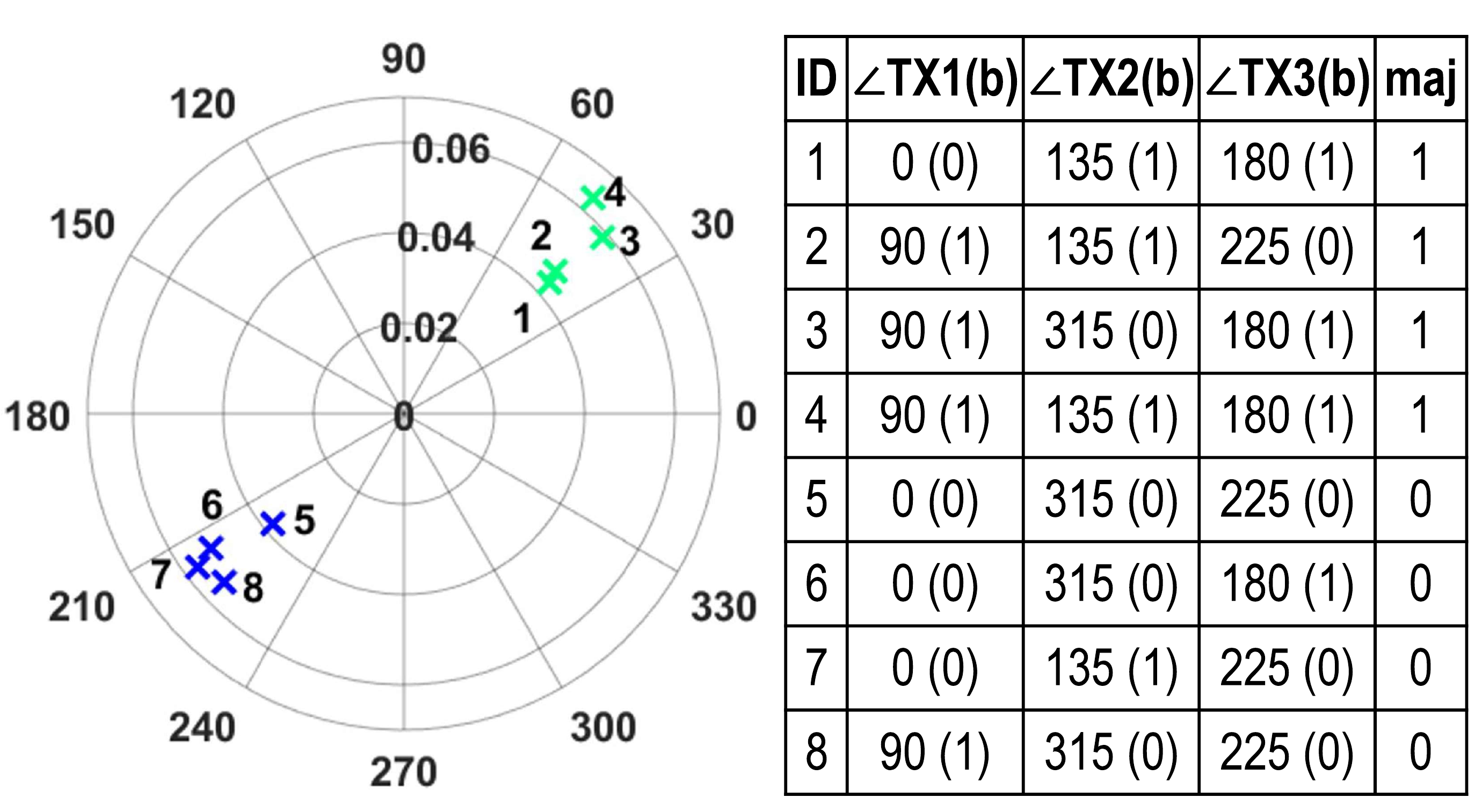}
    \vspace{-0.2cm}
    \caption{Constellation and truth table with transmitted phases/ bits for a specific RX. Blue/green symbols map to logical 0/1.}
    \label{fig:otaasig_tab2} 
    \vspace{-0.4cm}
\end{figure}

As an instance of the proposed approach and for illustration purposes, let us consider three TXs. In that case, we have a constellation with $2^3=8$ symbols for each RX. In order to map the eight symbols to their binary majority result, four corresponding to $maj(\cdot)=1$ and four corresponding to $maj(\cdot)=0$, decision regions are computed using the $K$-means clustering algorithm with $K=2$. We make sure that each cluster contains four symbols and that the combination of TX phases allows the mapping to the majority result. Fig.~\ref{fig:cst_res} shows an example of this method in three distinct RXs: on top, we show the received signals considering all possible bit combinations in the TXs and for all the swept phases, whereas, on bottom, we see the chosen constellations. Further, Fig.~\ref{fig:otaasig_tab2} shows the chosen transmitted phases for the case under study and how they are mapped in a particular receiver. %of the mapping. %and Table \ref{tab:majority_assign} show an example of the mapping.

\noindent
\textbf{Error rate assessment.} Once the candidate clusters are obtained, we compute the BER of each constellation in each RX, for all the different possible symbol phases, and choose the cluster that leads to the lowest average BER across RXs. In all cases, the BER has been evaluated considering the centroids of each binary cluster as ideal received symbols, and using the analytical expression of error rate of BPSK, 
\begin{equation}
    \label{eq:bpsk}
    %BER^{BPSK} = Q\bigg({\frac{0.5\cdot d_c}{\sqrt{N_0/2}}}\bigg),
    BER^{BPSK} = 0.5\cdot erfc\bigg({\frac{0.5\cdot d_c}{\sqrt{N_0}}}\bigg),
\end{equation}
where $erfc(\cdot)$ is the complementary error function, $d_c$ is the distance among centroids and $N_0$ is the noise spectral density. 

\noindent
\textbf{Bundling and accuracy evaluation.}
Once the final TX phases have been chosen considering the best average BER, an in-house Python HDC is used to evaluate its impact on the accuracy. Every associative memory connected to an RX stores 100 different prototype hypervectors, i.e., 100 different classes, each with 512-bit that suffices for the scenario considered in this paper. Errors coming from the OTA computations are modeled as uncorrelated bit flips over the query hypervectors.

While the baseline bundling consists on simply computing the bit-wise logical majority result across the different TX bits, we also consider a permuted bundling. This bundling consists on permuting the queries in the TXs prior to applying the majority operation to them. By permuting the hypervectors we obtain two benefits. First, this allows the identification of the transmitter of the detected class from the composite query. If we make each transmitter to apply a 1-bit cyclic permutation to its query before sending it to the wireless channel, the detected bundled hypervectors will contain the information of such permuted versions. Then, each receiver can expand its prototype hypervector set with their permuted versions, each corresponding to a different transmitter signature. The second direct benefit of permuting the hypervectors is that it helps increasing the quasi-orthogonality between them, which has a direct impact in accuracy, since the TXs share a common codebook of hypervectors.

\section{Results and Discussion} 
\label{results}
\begin{figure}[!t]
    \centering
    \vspace{-0.1cm}
    \includegraphics[width=0.9\columnwidth]{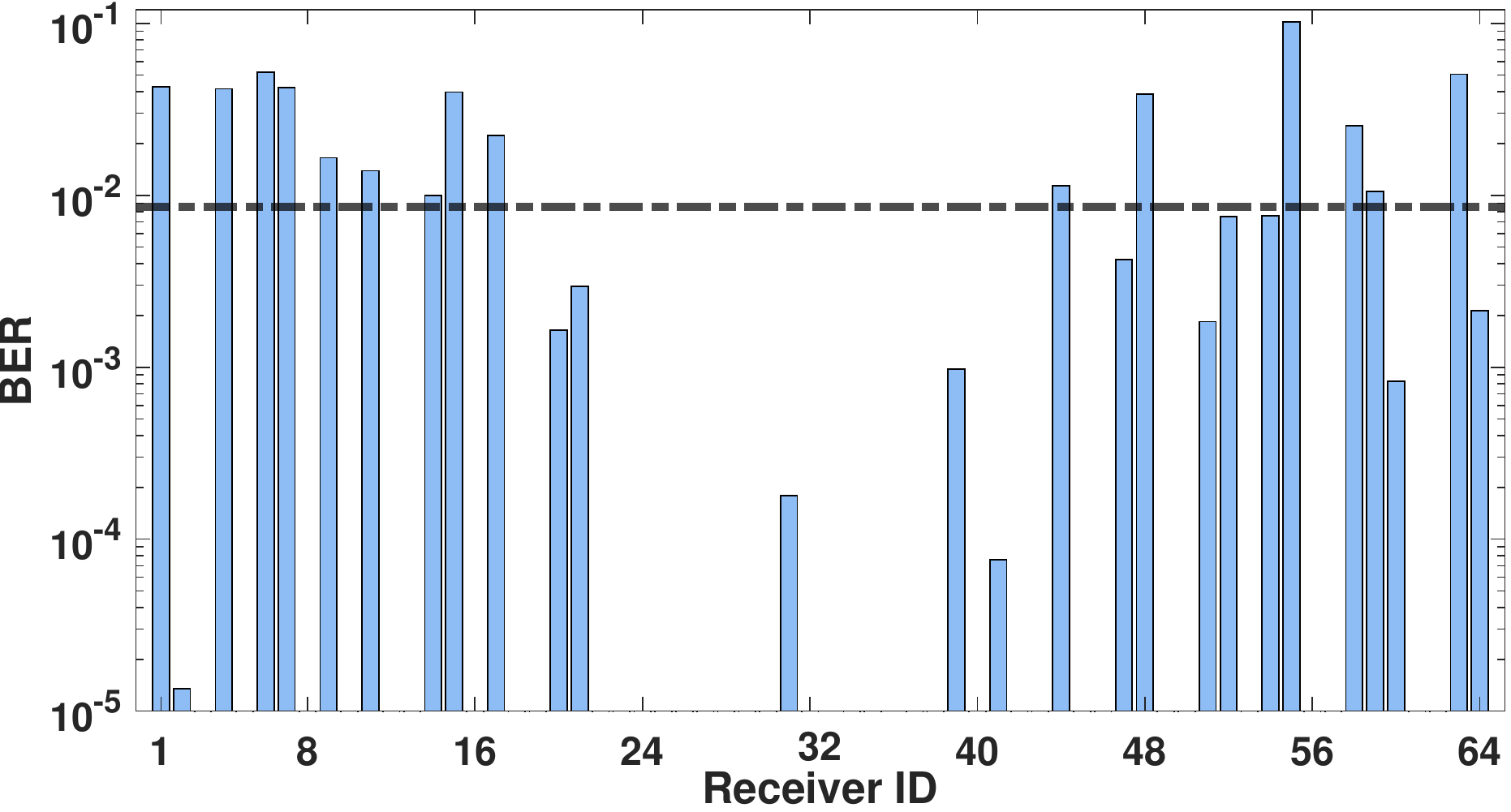}
    \caption{Resulting BER values per each individual RX in the architecture. The dashed line indicates the average value. }
    \label{fig:ber_Rxs} 
    \vspace{-0.3cm}
\end{figure}

%\hl{text here needs to be expanded}
%\textcolor{red}{Results from wireless BER and python HD framework}
After applying the proposed methodology and the careful optimization of the TX symbols as illustrated in Fig.~\ref{fig:cst_res}, we obtained the TX phases shown in Fig.~\ref{fig:otaasig_tab2} for our 3-TX system. The assessment of the error rate considering the chosen TX phases is summarized in Fig.~\ref{fig:ber_Rxs}, which plots the BER of each particular receiver in the 64-RX system under study. As it can be seen, the BER values are very much dependent on the particular receiver, with values lower than 10\textsuperscript{-5} in a significant amount of cases, but also with a worst-case BER of $\sim$0.1. In average, the error rate is below 0.01. Time-domain simulations, not shown for the sake of brevity, further confirm that the OTA computation can be done at multi-Gb/s rates.

%which according to Fig.~\ref{fig:ber_acc}, does not have any impact in the accuracy. Fig.~\ref{fig:ota_constellations} displays the constellations that are finally used in the architecture and how they map to the binary regions.

%the resulting link BERs are used to evaluate the accuracy in HDC. 

\begin{figure}[!t]
    \centering
    \vspace{-0.1cm}
    \includegraphics[width=0.9\columnwidth]{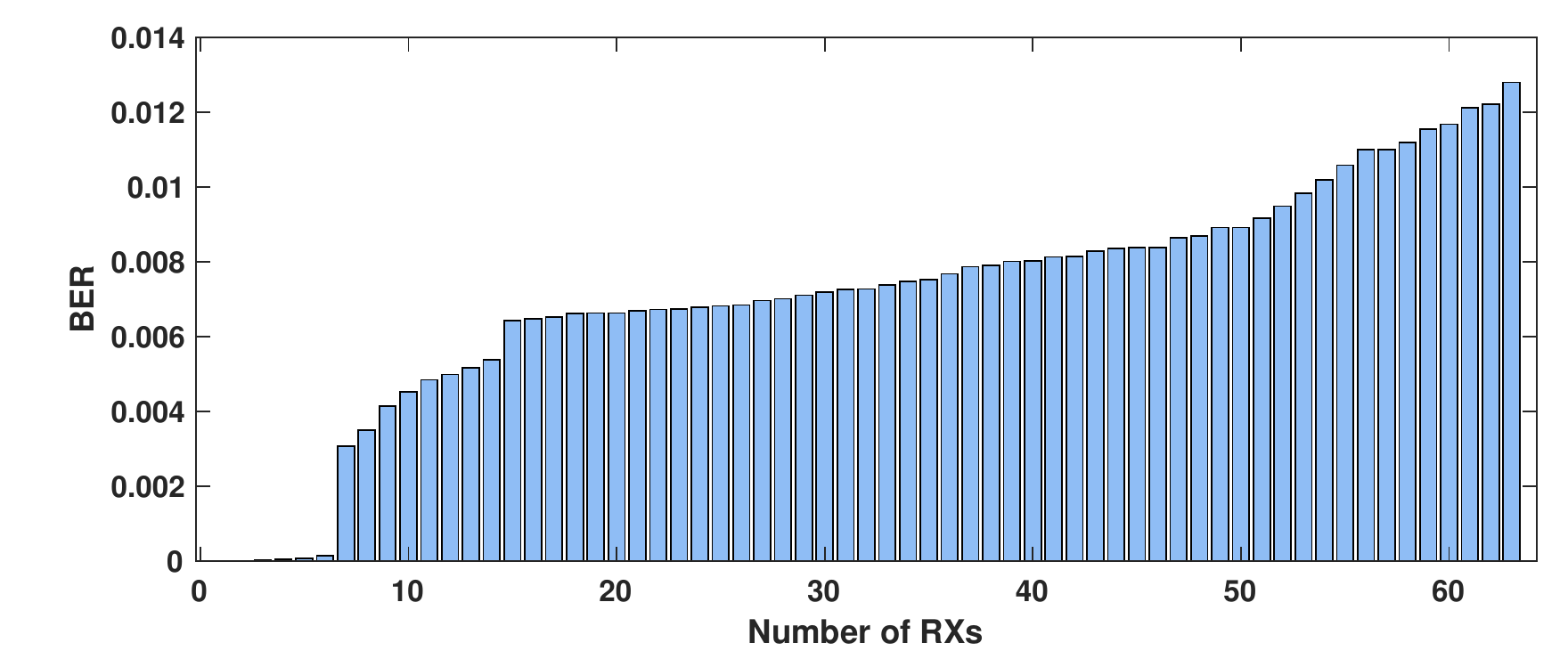}
    \caption{Architecture scalability in a 3 TXs scenario.}
    \label{fig:arch_scal} 
    \vspace{-0.3cm}
\end{figure}

To understand how the error rate could scale with the number of receivers, we re-simulate the entire architecture with a varying number of RX cores and computing the average BER obtained in each case. As shown in Fig.~\ref{fig:arch_scal}, the average BER generally increases with the number of receivers for which we are optimizing the architecture. This is expected since, when accommodating more constellations in our optimal TX phases search, we are imposing more conditions and hindering the joint optimization across all receivers.

% \begin{comment}
% \begin{figure} [!htb]
%     \centering
%     \vspace{-0.1cm}
%     \includegraphics[width=0.7\columnwidth]{images/cst_9c_res.png}
%     \caption{Phase sweep sample results}
%     \label{fig:cst_res} 
%     \vspace{-0.3cm}
% \end{figure}
% \begin{figure} [!htb]
%     \centering
%     \vspace{-0.1cm}
%     \includegraphics[width=1\columnwidth]{images/ota_nice.png}
%     \caption{Binary mapped constellations in the OTA case. Blue symbols map to 0 and
% green symbols map to 1. \textcolor{red}{REDO or remove}}
%     \label{fig:ota_constellations} 
%     \vspace{-0.3cm}
% \end{figure}
% \end{comment}

\begin{figure}[!t]
    \centering
    \includegraphics[width=1\columnwidth]{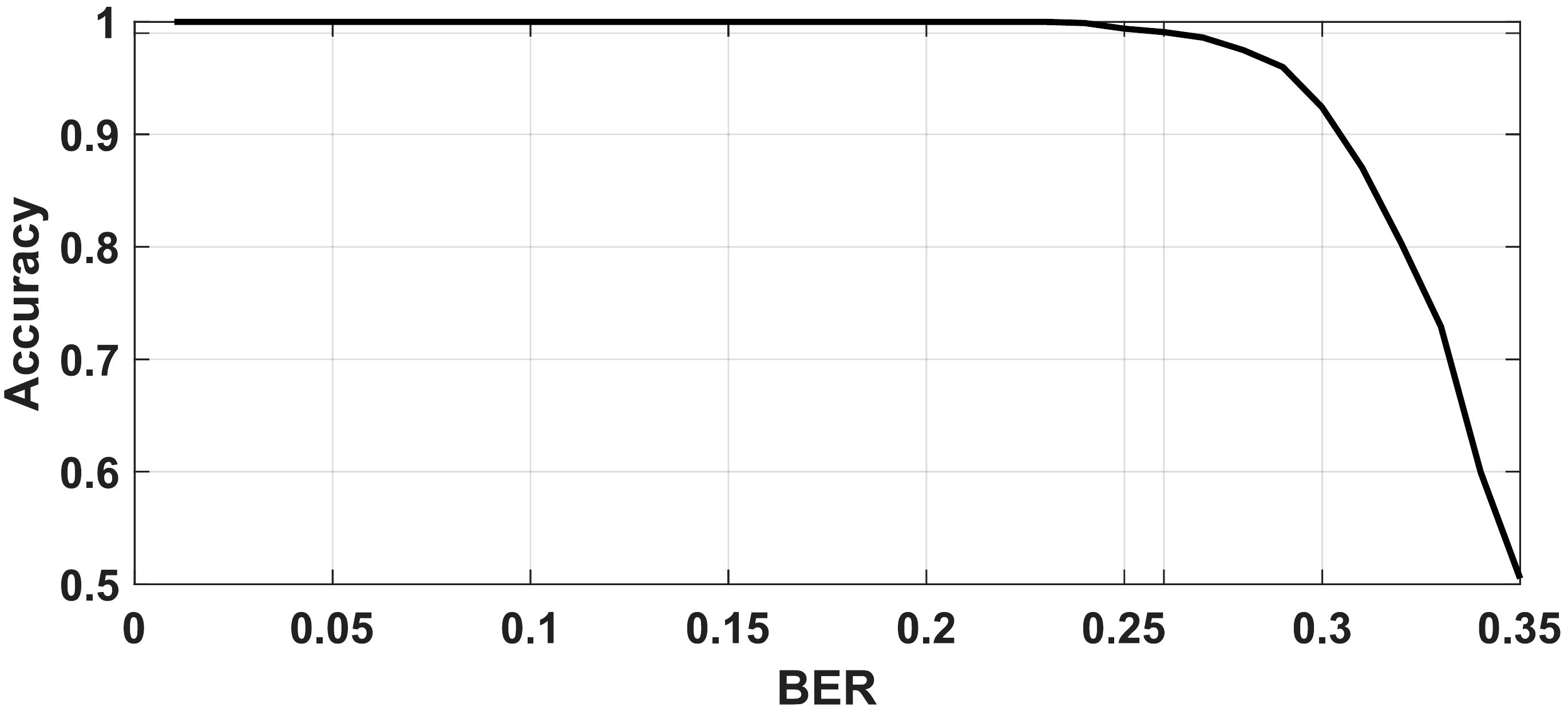}
    \vspace{-0.5cm}
    \caption{Impact on the accuracy of a classification task when increasing the error rate of the encoder-to-search interconnect.}
    \label{fig:ber_acc} 
    \vspace{-0.3cm}
\end{figure}

%%%%%%%% SIMIILARITY RESULTS %%%%%%%%%%
Next, to evaluate the performance of the proposed architecture, we execute a typical HDC-based classification task by introducing the wireless error figures in the HDC chain. First, we illustrate the impact of errors on the classification by performing a generic classification task test over 100 prototype hypervectors of 512 bits, with increasing error rates. As Fig.~\ref{fig:ber_acc} depicts, the class accuracy remains above 99\% even when we apply bit flips equivalent to a BER of 0.26. This means that the noise robustness provided by the HDC properties relaxes the error link conditions, ensuring a correct behaviour under the worst-case wireless scenarios, as we show next.

\begin{figure}[!t]
    %\centering
    \begin{subfigure}{0.49\textwidth}
    \includegraphics[width=1\textwidth]{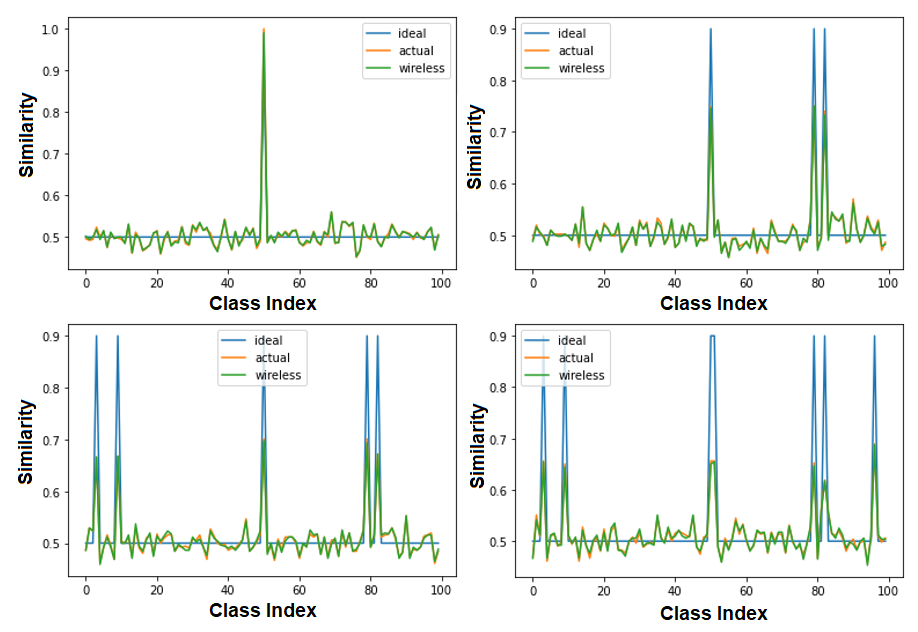}
    \vspace{-0.5cm}
    \caption{Baseline bundling}
    \label{fig:sim_res_baseline} 
    \end{subfigure}
    \begin{subfigure}{0.5\textwidth}
    %\centering
    \includegraphics[width=1\textwidth]{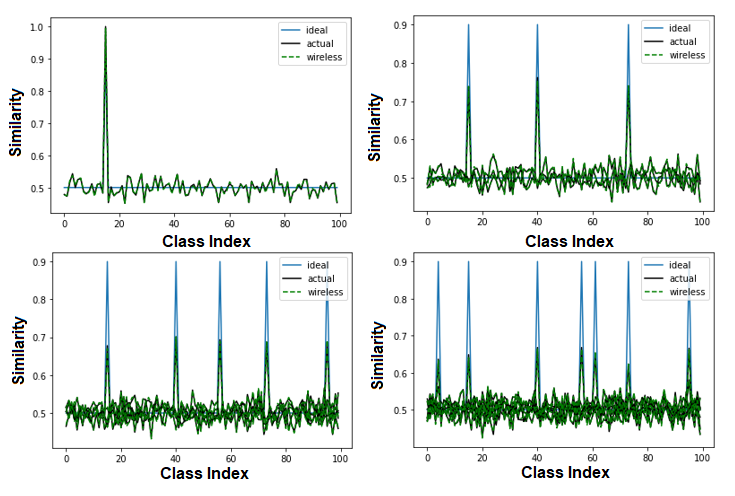}
    \vspace{-0.5cm}
    \caption{Permuted bundling}
    \label{fig:sim_res_permuted} 
    \end{subfigure}
    \caption{Similarity results comparison for different forms of bundling and number of bundled hypervectors. We show bundling of one, three, five and seven hypervectors.}
    \vspace{-0.2cm}
\end{figure}

Fig.~\ref{fig:sim_res_baseline} and Fig.~\ref{fig:sim_res_permuted} show the similarity search result for the baseline bundling and permuted bundling cases, respectively, after comparing the composite query hypervector against a set of 100 prototype hypervectors. The figures show how a single query has capacity enough to successfully accommodate several queries via bundling (blue line), and that the error introduced by the wireless OTA computation reduces the similarity but does not introduce any classification errors (green line). Table~\ref{tab:hdframe1} shows the numerical results of the final class accuracy for the executed task, comparing an ideal channel without errors with our wireless channel with a sizable BER. The effect of the wireless channel is practically irrelevant in terms of accuracy, as predicted by Fig.~\ref{fig:ber_acc}. Moreover, the permuted bundling significantly improves the baseline bundling, confirming that the proposed approach supports the aggregation of a dozen hypervectors over the air and the parallelization of similarity search over tens of IMCs. 

\begin{table}[!t]
\caption{Accuracy results in an IMC for the analyzed bundling techniques, a variable number of TXs, both for an ideal channel (no errors) and for a channel with BER equivalent, in average, to that obtained with 64 RXs.}
\vspace{-0.1cm}
\centering
\label{tab:hdframe1}
\begin{tabular}{|c|c|c|c|c|c|c|c|}
\hline 
\multirow{4}{*}{\shortstack{Baseline\\Bundling}} & & \multicolumn{6}{c|}{\textbf{Number of bundled hypervectors}} \\ \cline{3-8}
 & \textbf{Channel} & 1 & 3 & 5 & 7 & 9 & 11\\
\cline{2-8}
 & Ideal & 1 & 0.966 & 0.902 & 0.803 & 0.704 & 0.543\\
\cline{2-8}
 & Wireless & 1 & 0.966 & 0.9 & 0.801 & 0.699 & 0.537 \\ 
\hline \hline
\multirow{4}{*}{\shortstack{Permuted\\Bundling}} & & \multicolumn{6}{c|}{\textbf{Number of bundled hypervectors}} \\ \cline{3-8}
 & \textbf{Channel} & 1 & 3 & 5 & 7 & 9 & 11\\
\cline{2-8}
 & Ideal & 1 & 1 & 1 & 1 & 0.995 & 0.978 \\
\cline{2-8}
 & Wireless & 1 & 1 & 1 & 1 & 0.994 & 0.963\\ 
\hline
\end{tabular} 
\vspace{-0.3cm}
\end{table}

% \begin{table}[h]
% \caption{Accuracy results in an IMC with baseline bundling for a different number of TXs and a BER equivalent to the 64 RXs average.}
% \centering
% \label{tab:hdframe1}
% \begin{tabular}{|c|c|c|c|c|c|c|}
% \hline 
% & \multicolumn{6}{|c|}{\textbf{Number of bundled hypervectors}} \\
% \textbf{Channel} & 1 & 3 & 5 & 7 & 9 & 11\\
% \hline
% Ideal & 1 & 0.966 & 0.902 & 0.803 & 0.704 & 0.543\\
% \hline
% Wireless & 1 & 0.966 & 0.9 & 0.801 & 0.699 & 0.537 \\
% \hline
% \end{tabular} 
% \end{table}

% \begin{table}[h]
% \caption{Accuracy results in an IMC with permuted bundling for a different number of TXs and a BER equivalent to the 64 RXs average.}
% \centering
% \begin{tabular}{|c|c|c|c|c|c|c|}
% \hline 
% & \multicolumn{6}{|c|}{\textbf{Number of bundled hypervectors}} \\
% \textbf{Channel} & 1 & 3 & 5 & 7 & 9 & 11 \\
% \hline
% Ideal & 1 & 1 & 1 & 1 & 0.995 & 0.978 \\
% \hline
% Wireless & 1 & 1 & 1 & 1 & 0.994 & 0.963\\
% \hline
% \end{tabular}

% \label{tab:hdframe2}
% \end{table}

%\begin{figure} [!htb]
%    \centering
%    \vspace{-0.1cm}
%    \includegraphics[width=1\columnwidth]{images/hdframe5}
%    \caption{Similarity results multibit bundling}
%    \label{fig:sim_res_multibit} 
%    \vspace{-0.3cm}
%\end{figure}

%\begin{figure} [!htb]
%    \centering
%    \vspace{-0.1cm}
%    \includegraphics[width=1\columnwidth]{images/hdframe4}
%    \caption{Similarity results MANN. Reduce this.}
%    \label{fig:sim_res_mann} 
%    \vspace{-0.3cm}
%\end{figure}

\section{Conclusion} \label{cncl}
\vspace{-0.1cm}

In this work, we introduced an OTA on-chip computing concept capable of overcoming the scalability bottleneck present in wired NoC architectures when scaling out IMC-based HDC systems. By using a WNoC communication layer, a number of encoders is able to concurrently brodacast HDC queries towards all the IMC cores within the architecture. Then, a pre-characterization of the propagation environment allows to map the received constellations to the computed composite query, in each core, based on a decision region strategy. Through a proper correspondence between the TX phases, the received constellation and the decision region, we have shown that the opportunistic calculation of the bit-wise majority of the transmitted HDC queries is possible with low error. We demonstrated the concept and shown its scalability up to 11 TXs and 64 RXs, obtaining the BER of the OTA approach and later employing it to evaluate the impact of the WNoC errors in a HDC classification task. Overall, we conclude that the quality of the WNoC links are solid enough to have a negligible impact on the application accuracy, mostly thanks to the great error robustness of HDC. 

%Moreover, we test the effects of different query bundling and show that a permuted bundling has a positive impact in accuracy.

%non-trivial communications
%\input{07-Ack}

%\appendices

%\section{}

\section*{Acknowledgment}
Authors gratefully acknowledge funding from the European Union’s Horizon 2020 research and innovation programme under grant agreement No 863337 (WiPLASH).

%Huawei?

\footnotesize
\bibliographystyle{IEEEtran}
\tiny
\bibliography{bstctl,ref}
\end{document}